\documentclass[twocolumn,showpacs,preprintnumbers,amsmath,amssymb]{revtex4}

\usepackage{graphicx}
\usepackage{dcolumn}
\usepackage{bm}
\usepackage{color}

\begin{document}

\preprint{}

\title{Dynamical mean-field theory for quantum spin systems:\\ Test of solutions for magnetically ordered states}

\author{Junya Otsuki$^{1,2}$}
\author{Yoshio Kuramoto$^{2}$}
\affiliation{%
$^1$Theoretical Physics III, Center for Electronic Correlations and Magnetism,\\
Institute of Physics, University of Augsburg, D-86135 Augsburg, Germany\\
$^2$Department of Physics, Tohoku University, Sendai 980-8578, Japan
}%

\date{\today}

\begin{abstract}
A spin version of 
dynamical mean-field theory is extended for magnetically ordered states
in the Heisenberg model. 
The self-consistency equations are solved
with high numerical accuracy by means of the continuous-time quantum Monte Carlo 
with bosonic baths coupled to the spin.
The resultant solution is critically tested by known physical properties.
In contrast with the mean-field theory,  soft paramagnons appear near the transition temperature. 
Moreover, the Nambu-Goldstone mode (magnon)
in the ferromagnetic phase is reproduced reasonably well.
However, antiferromagnetic magnons have an energy gap
in contradiction to the Nambu-Goldstone theorem.
The origin of this failure is 
discussed in connection with an artificial first-order nature of the transition.
\end{abstract}



\maketitle

\section{Introduction}

Approaches from high dimensions are effective ways for addressing magnetism in quantum spin systems.
The molecular-field (MF) approximation becomes exact in the limit $z \to \infty$,
where $z$ denotes the number of interacting neighbors.
In order to address three-dimensional systems, 
multitudes of extensions have been developed to take account of fluctuations around the MF.
A natural direction is to include the leading correction of $1/z$
in a self-consistent fashion.  This corresponds to going one step beyond the mean-field theory. 
For classical spin systems, such extension was proposed long-time ago\cite{Brout60} in the name of the spherical model approximation. 
The quantum version of the theory 
has been constructed more recently,\cite{Kuramoto-Fukushima98}
which may be regarded as 
a spin version of 
the dynamical mean-field theory (DMFT).\cite{Georges96} 
The spin DMFT (s-DMFT) is closely related to theories for quantum spin glass in high dimensions\cite{Bray-Moore80, Sachdev-Ye93, Grempel98, Georges00}
and impurities in antiferromagnet.\cite{Vojta00}


For electron systems,
spin and charge fluctuations 
beyond the mean-field level has been considered in a scheme called the extended DMFT.\cite{Smith-Si00, Haule02, Sun-Kotliar02}  The level of approximation for spin fluctuations is the same as the original s-DMFT.\cite{Kuramoto-Fukushima98}
Further improvement to include nonlocal correlations has been developed 
recently.\cite{Rubtsov12, Ayral13}
These theories provide systematic ways of incorporating intersite interactions 
beyond the s-DMFT or the extended DMFT.
In order to construct reliable yet practical framework in this direction, 
it is desirable to clarify the significance and limitation of 
the approximation first in pure spin systems.

Because of numerical complexity of the s-DMFT, however, 
explicit solution 
of the self-consistent equations
has not been available.
Instead, an additional approximation such as a static approximation\cite{Fukushima-Kuramoto98} has been adopted.
Furthermore, account of a long-range order has not been worked out so far.
In this paper, we present numerical solution of the fully
self-consistent equations for both paramagnetic (PM) and magnetically ordered states by using a recently developed numerical method\cite{Otsuki-SB} based on the continuous-time quantum Mote Carlo method (CT-QMC).\cite{Gull11}
We shall check the solutions against some known properties, e.g., emergence of gapless excitations in ordered states.

In the next section, we review the formalism with particular care of its application to ferromagnetic (FM) and antiferromagnetic (AFM) states.
Numerical solutions are presented in Sec.~\ref{sec:results}.
In particular, it will be demonstrated that a fictitious energy gap arises in magnetic excitations from the antiferromagnetic state.
We shall discuss the origin of this artifact in detail in Sec.~\ref{sec:discussion}.
Finally, a summary is given in Sec.~\ref{sec:summary}.

\section{Formalism}

\subsection{Self-consistent equations}
\label{sec:self_consist}

We consider $1/z$ corrections around the MF in terms of the scaling $J \propto 1/z$ of the exchange coupling constant.
This scaling keeps the transition temperature finite in the limit $z \to \infty$ within the MF approximation, 
$T_{\rm c}^{\rm MF} \sim z J \sim O(1)$, 
and allows discussion of magnetic phase transitions.
Fluctuations are taken into account as corrections of order $1/z$ and higher for the polarization function $\Pi_{\bm{q}}(i\nu_n)$ (irreducible part of the susceptibility).
It has been demonstrated\cite{Kuramoto-Fukushima98} that the momentum-dependence of $\Pi_{\bm{q}}(i\nu_n)$ first appears as a $1/z^2$ correction, meaning that $\Pi_{\bm{q}}(i\nu_n)$ is local in space but non-local in time within $O(1/z)$: $\Pi_{\bm{q}}(i\nu_n) \to \Pi(i\nu_n)$.
Hence, the s-DMFT, which takes full account of dynamical local fluctuations for $\Pi(i\nu_n)$, provides a solution being valid within $O(1/z)$.
The dynamical fluctuations incorporated in the s-DMFT include not only the $1/z$ corrections but also higher-order terms within local approximation.

Here, we mention another scaling $J \propto 1/\sqrt{z}$ 
which is used in some literatures.\cite{Smith-Si00, Haule02}
This scaling gives a divergent transition temperature
$T_{\rm c}^{\rm MF}\propto \sqrt{z}$ in the Heisenberg model. 
However it is useful in 
a disordered system where the coupling constants $J_{ij}$ distribute between positive and negative values.\cite{Bray-Moore80, Sachdev-Ye93, Grempel98, Georges00, Parcollet99, Otsuki-Vollhardt}
In this case, the MF term averages out to suppress the long-range order, and fluctuations remain in the limit $z\to \infty$.

There are several ways of deriving the effective local model as in the DMFT.\cite{Georges96}
The basic idea is that one introduces an auxiliary field 
which locally couples to the spin.
Then one applies the DMFT to this field.
Thus, dynamical fluctuations of the order parameter
is taken into account within the local approximation.
An important difference to the ordinary DMFT is that the auxiliary field may have a finite expectation value corresponding to the MF.
This situation is similar to the treatment of the Bose condensation in the bosonic DMFT.\cite{Byczuk-Vollhardt08, Anders11}
In the following, we first review the equations derived in Ref.~\cite{Kuramoto-Fukushima98} and then consider applications to the FM and AFM states.

We consider the $S=1/2$ Heisenberg model with nearest-neighbor interactions on a $d$-dimensional hyper-cubic lattice ($z=2d$).
The Hamiltonian reads
\begin{align}
\label{eq:hamil}
H = 
- \frac{J}{2} \sum_{\langle i j \rangle} \bm{S}_i \cdot \bm{S}_j. 
\end{align}
It is convenient to work in the path integral formalism.
We use a pseudo-fermion representation
$\bm{S}_i 
= (1/2) \sum_{\sigma \sigma'} 
c_{i\sigma}^{\dag} \bm{\sigma}_{\sigma \sigma'} c_{i\sigma'}$, 
with $\bm{\sigma}$ being the Pauli matrix.
The constraint $\sum_{\sigma} c_{i\sigma}^{\dag} c_{i\sigma} = 1$
is imposed by the Popov-Fedotov method.\cite{Popov-Fedotov88}
The partition function $Z$ is thus expressed as
\begin{align}
&Z = 
\int \left( \prod_i {\cal D}\bm{S}_i \right) 
\nonumber \\
&\times \exp \Bigg[
-\sum_i {\cal S}_{{\rm B}i}
+\int_0^{\beta} d\tau 
\frac{J}{2} \sum_{\langle i j \rangle}
 \bm{S}_i(\tau) \cdot \bm{S}_j(\tau)
\Bigg].
\end{align}
Here, the spin path integral including the Berry phase term ${\cal S}_{{\rm B}i}$ is defined by
\begin{align} 
\int &{\cal D}\bm{S}_i 
e^{-{\cal S}_{{\rm B}i}} 
\equiv i^N
\int \prod_{\sigma} {\cal D}c_{i\sigma}^{*} {\cal D}c_{i\sigma}
\nonumber \\
&\times \exp \left[
-\sum_{\sigma} \int_0^{\beta} d\tau
c_{i\sigma}^{*}(\tau) (\partial_{\tau} +i\pi/2\beta) c_{i\sigma}(\tau)
\right],
\end{align}
where $c_{i\sigma}(\tau)$ and $c_{i\sigma}^{*}(\tau)$ are Grassmann numbers
and $N$ denotes the number of sites.
Because of the imaginary chemical potential $\mu=-i\pi/2\beta$,
contributions from the empty and doubly occupied states cancel out to eliminate charge fluctuations.\cite{Popov-Fedotov88}

We define the dynamical susceptibility 
\begin{align}
\chi^{\xi \xi'}_{\bm{q}}(i\nu_n)
=\int_0^{\beta} d\tau\langle S^{\xi}_{\bm{q}}(\tau) S^{\xi'}_{-\bm{q}} \rangle
e^{i\nu_n \tau}
\equiv
\langle S^{\xi}_{\bm{q}}; S^{\xi'}_{-\bm{q}} \rangle,
\label{eq:suscep}
\end{align}
with $\xi=x, y, z$
and 
$\nu_n = 2\pi nT$.
Hereafter, $\hat{\chi}_{\bm{q}}$ denotes $3 \times 3$ matrix with respect to $\xi$ and $\xi'$.
In a local approximation, 
$\hat{\chi}_{\bm{q}} (i\nu_n)$ is expressed in terms of the polarization function $\hat{\Pi}(i\nu_n)$ as
\begin{align}
\label{eq:chi_q} 
\hat{\chi}_{\bm{q}}^{-1} (i\nu_n)
= \hat{\Pi}^{-1}(i\nu_n)
- J_{\bm{q}},
\end{align}
where
$J_{\bm{q}}=zJ \gamma_{\bm{q}}$ with 
$\gamma_{\bm{q}}=d^{-1} \sum_{i=1}^{d} \cos q_{i}$.

Instead of solving the lattice problem directly,
we solve
an alternative effective local problem with a dynamical interaction $\hat{\cal J}(i\nu_n)$ which incorporates dynamics of interacting spins.
The action ${\cal S}_{\rm loc}$ at arbitrary site 
is given by
\begin{align}
{\cal S}_{\rm loc}
= {\cal S}_{{\rm B}i}
-\frac{1}{2} \int d\tau d\tau'
\Delta \bm{S}_i^{\rm t} (\tau)
\hat{\cal J}(\tau-\tau')
\Delta \bm{S}_i (\tau')
\nonumber \\
- \bar{\bm{h}}_i
\cdot 
\int d\tau
\bm{S}_i(\tau),
\label{eq:S_loc}
\end{align}
where 
$\Delta \bm{S}_i$ is defined by
$\Delta \bm{S}_i = \bm{S}_i - \langle \bm{S}_i \rangle$ and 
$\bar{\bm{h}}_i=\sum_j J_{ij} \langle \bm{S}_j \rangle$ 
is the MF.
From the dynamical susceptibility $\chi_{\rm loc}(i\nu_n)$ evaluated with this local model, $\hat{\Pi}(i\nu_n)$ is obtained as
\begin{align} 
\hat{\chi}_{\rm loc}^{-1}(i\nu_n)
= \hat{\Pi}^{-1}(i\nu_n)
- \hat{\cal J}(i\nu_n).
\label{eq:chi_loc}
\end{align}
The self-consistency condition reads that the local susceptibility $\chi_{\rm loc}$ coincides with the site-diagonal component of the susceptibility in the original lattice model, i.e.,
\begin{align}
\hat{\chi}_{\rm loc}(i\nu_n) 
= \langle \hat{\chi}_{\bm{q}}(i\nu_n) \rangle_{\bm{q}},
\label{eq:self_consist}
\end{align}
where 
$\langle \cdots \rangle_{\bm{q}} = N^{-1} \sum_{\bm{q}} (\cdots)$.
Thus, $\hat{\cal J}(i\nu_n)$ and $\hat{\Pi}(i\nu_n)$ are
determined in a self-consistent manner.

From the self-consistency condition, we can demonstrate that 
the real part of the diagonal elements of $\hat{\cal J}(\tau)$ is positive.
Hence the interaction in ${\cal S}_{\rm loc}$ stabilizes the magnetic moment.

\subsection{Effective Hamiltonian for paramagnetic state}
We present the impurity Hamiltonian describing the effective local problem.
The CT-QMC, which we shall use as the impurity solver, does not require the Hamiltonian.
However it is necessary if one uses, for example, the exact diagonalization method. 
The frequency-dependent interaction ${\cal J}(i\nu_n)$ 
can be expressed in terms of 
a vector bosonic field $\bm{b}_{\bm{q}}$ which mediates the local retarded interaction.
In the PM phase without magnetic field, 
the Hamiltonian is written as
\begin{align} 
H_{\rm imp}^{\rm PM}
= \sum_{\bm{q}} \omega_{\bm{q}} 
\bm{b}_{\bm{q}}^{\dag}
\cdot
\bm{b}_{\bm{q}}
+ g \bm{S} \cdot
(\bm{b} +\bm{b}^{\dag}),
\label{eq:H_imp-PM}
\end{align}
where $\bm{b}=N^{-1/2} \sum_{\bm{q}} \bm{b}_{\bm{q}}$.
The propagator of the bosons describes the effective interaction $\hat{\cal J}(i\nu_n)$,
which is now scalar and given by
\begin{align}
\label{eq:J_eff1}
{\cal J}(i\nu_n)
= \frac{g^2}{N} \sum_{\bm{q}}
\frac{2\omega_{\bm{q}}}{\nu_n^2 + \omega_{\bm{q}}^2}.
\end{align}

The model (\ref{eq:H_imp-PM}) 
may be referred to as the bosonic Kondo model
or the SU(2) spin-boson model, and
has been investigated in some literatures.\cite{Vojta00, Otsuki-SB, Zhu-Si02, Zarand-Demler02, Guo12}
In a particular situation, the model exhibits an intermediate-coupling fixed point where bosonic fluctuations lead to a critical behavior.
But we should emphasize that the spin entropy is not quenched within the impurity level in contrast to the ordinary (fermionic) Kondo model.\cite{unpublished}
Hence, a long-range order is expected as self-consistent solution
unless it is prohibited by low-dimensionality.

\subsection{Ferromagnetic state}
We should take care of the off-diagonal element of $\hat{\cal J}$
in magnetic phases or under magnetic field.
To make the discussion concrete, 
we restrict ourselves to a situation where the magnetic moment is parallel to $z$ axis:
$\langle S^x \rangle = \langle S^y \rangle = 0$,
$\langle S^z \rangle \neq 0$.
In this situation, $\chi^{xy}$ is non-zero,
and it is convenient to use basis which diagonalizes $\hat{\chi}$.
From the symmetry relations $\chi^{xx} = \chi^{yy}$ and 
$\chi^{xy}=-\chi^{yx}$,
we obtain eigenvalues 
$\chi^{xx} -i \chi^{xy} \equiv \chi^{\perp}$,
$\chi^{xx} +i \chi^{xy}$ and
$\chi^{zz} \equiv \chi^{\parallel}$.
Here, $\chi^{\perp}$ corresponds to the correlation 
$\chi^{\perp}(i\nu_n)=\langle S^+; S^{-} \rangle/2$,
and the second eigenvalue equals to
$\chi^{\perp}(i\nu_n)^*$ in the Matsubara-frequency domain.
Using this representation, 
we may write $S_{\rm loc}$ in Eq.~(\ref{eq:S_loc}) as
\begin{align}
{\cal S}_{\rm loc}
= {\cal S}_{{\rm B}i}
-\frac{1}{2} \int d\tau d\tau' \Big[
S^z_i (\tau)
{\cal J}^{\parallel}(\tau-\tau')
S^z_i (\tau')
\nonumber \\
+S^{-}_i (\tau)
{\cal J}^{\perp}(\tau-\tau')
S^{+}_i (\tau') \Big]
\nonumber \\
-\left[ \bar{h}_i - {\cal J}^{\parallel}(0) \langle S^z_i \rangle \right]
\int d\tau S^z_i(\tau),
\label{eq:S_loc2}
\end{align}
where ${\cal J}^{\parallel}(0)$ is the static part, which is referred to as the reaction field.

The Hamiltonian for the magnetic state is more complicated than that for the PM state, Eq.~(\ref{eq:H_imp-PM}).
The finite expectation value of $\langle S^z \rangle$ yields the high-frequency behavior
\begin{align}
{\cal J}^{\perp}(i\nu_n) 
\sim c \chi^{\perp}_{\rm loc}(i\nu_n) 
\sim -c \langle S^z \rangle /i\nu_n,
\label{eq:high_freq}
\end{align}
with $c=\langle J_{\bm{q}}^2 \rangle_{\bm{q}}$, 
while ${\cal J}(i\nu_n)$ in Eq.~(\ref{eq:J_eff1}) is proportional to $1/(i\nu_n)^2$.
The $1/i\nu_n$-term in $\chi_{\rm loc}^{\perp}$ is due to the commutation relation of the spin operators, 
$[S^{+}, S^{-}] = 2S^z$, 
while its absence in ${\cal J}(i\nu_n)$ is 
due to commutativity of the displacement operators
$\bm{\phi}=\bm{b}+\bm{b}^{\dag}$, i.e., $[\phi^{\xi}, \phi^{\xi'}]=0$.
This consideration leads to an additional coupling between the spin and the canonical momentum 
$\bm{\pi} = (\bm{b} - \bm{b}^{\dag})/i$.
We thus arrive at the Hamiltonian which describes the effective impurity model in the FM phase:
\begin{align}
&H_{\rm imp}^{\rm FM}
= \sum_{\bm{q}} \sum_{\eta=z,+,-} \omega_{\bm{q} \eta} 
b_{\bm{q} \eta}^{\dag}
b_{\bm{q} \eta}
- \left[ \bar{h} - {\cal J}^{\parallel}(0) \langle S^z \rangle \right] S^z
\nonumber \\
&+ g_{z} S^z \phi^z
+ g_{\perp} \left[ S^{+} (\alpha \phi^{-} + i\beta \pi^{-})
+S^{-} (\alpha \phi^{+} - i\beta \pi^{+}) \right],
\label{eq:H_imp-FM}
\end{align}
where 
the operators for $xy$ components are transformed 
into the basis which diagonalizes the susceptibility:
\begin{align}
&b_{\bm{q}\pm}= (b_{\bm{q}x} \pm i b_{\bm{q}y})/\sqrt{2},
\nonumber \\
&b_{\bm{q}\pm}^{\dag}= (b_{\bm{q}x}^{\dag} \mp i b_{\bm{q}y}^{\dag})/\sqrt{2},
\nonumber \\
&\phi^{\pm} 
= (\phi^x \pm i\phi^y )/\sqrt{2}
= b_{\pm} + b_{\mp}^{\dag},
\nonumber \\
&\pi^{\pm} 
= (\pi^x \pm i\pi^y )/\sqrt{2}
= -i (b_{\pm} - b_{\mp}^{\dag}).
\label{eq:op_perp}
\end{align}
The latter two operators follow the commutation relation
$[\phi^{\pm}, \pi^{\mp}]=2i$.
We parameterize $\alpha$ and $\beta$ as
$\alpha = \cos (\theta/2)$ and 
$\beta = \sin (\theta/2)$ and 
$\theta$ is determined later.
This Hamiltonian corresponds to the action ${\cal S}_{\rm loc}$ in Eq.~(\ref{eq:S_loc2}) where the effective interaction is given by (see Appendix~\ref{app:J_eff} for detail)
\begin{align}
{\cal J}^{\perp}(i\nu_n) 
= -\frac{g_{\perp}^2}{N} \sum_{\bm{q}}
\left( 
\frac{1-\sin \theta}{i\nu_n - \omega_{\bm{q}+}}
+\frac{1+\sin \theta}{-i\nu_n - \omega_{\bm{q}-}}
\right).
\label{eq:J_eff2}
\end{align}
On the other hand,
${\cal J}^{\parallel}(i\nu_n)$ is given by Eq.~(\ref{eq:J_eff1}) with replacing $g$ and $\omega_{\bm{q}}$ by $g_{z}$ and $\omega_{\bm{q}z}$, respectively.
Because of the non-commutativity
between $\phi^{\pm}$ and $\pi^{\mp}$, 
${\cal J}^{\perp}(i\nu_n)$ decays in proportion to $1/i\nu_n$.
Comparing with Eq.~(\ref{eq:high_freq}), 
we obtain
$2 g_{\perp}^2 \sin \theta= -c \langle S^z \rangle$.
We can confirm that
the above expressions reduce to those for the PM state,
when $\langle S^z \rangle=0$ and 
$\omega_{\bm{q}+}=\omega_{\bm{q}-}$.

\subsection{Antiferromagnetic state}
The AFM state can be addressed in terms of two-sublattice as in DMFT.
We introduce two polarization functions $\Pi_{\rm A}^{\lambda}(i\nu_n)$ and $\Pi_{\rm B}^{\lambda}(i\nu_n)$ for each sublattice named A and B 
with $\lambda =\ \parallel, \perp$.
Then, the susceptibility $\tilde{\chi}_{\bm{q}}^{\lambda}$ in the reduced Brillouin zone (RBZ) is given by
\begin{align}
\tilde{\chi}_{\bm{q}}^{\lambda}(i\nu_n) 
= \begin{pmatrix}
\Pi_{\rm A}^{\lambda}(i\nu_n)^{-1} & -J_{\bm{q}} \\
-J_{\bm{q}} & \Pi_{\rm B}^{\lambda}(i\nu_n)^{-1}
\label{eq:chi_q-AFM}
\end{pmatrix}^{-1}.
\end{align}
Here, we have used the relation
$J_{\bm{q}}=-J_{\bm{q}+\bm{Q}}$ with $\bm{Q}=(\pi, \cdots, \pi)$,
which holds for nearest-neighbor interactions on bipartite lattices.
The site-diagonal component 
\begin{align}
\chi_{\alpha}^{\lambda} 
= \frac{2}{N} \sum_{\bm{q} \in \text{RBZ}} 
\langle \alpha | \tilde{\chi}_{\bm{q}}^{\lambda} | \alpha \rangle
\end{align}
at sublattices $\alpha={\rm A}, {\rm B}$
is given in the original Brillouin zone by
\begin{align}
\chi_{\alpha}^{\lambda}(i\nu_n)
=\frac{1}{N} \sum_{\bm{q}}
\frac{\Pi_{\bar{\alpha}}^{\lambda}(i\nu_n)^{-1}}{\Pi_{\alpha}^{\lambda}(i\nu_n)^{-1} \Pi_{\bar{\alpha}}^{\lambda}(i\nu_n)^{-1} - J_{\bm{q}}^2},
\label{eq:chi_diag-AFM}
\end{align}
with $\bar{\alpha}$ being $\bar{\rm A}={\rm B}$ and $\bar{\rm B}={\rm A}$.
This expression reduces to 
$\langle \chi_{\bm{q}} (i\nu_n) \rangle_{\bm{q}}$ 
in Eq.~(\ref{eq:self_consist}) 
when $\Pi_{\rm A}^{\lambda}=\Pi_{\rm B}^{\lambda}$, 
by the relation $J_{\bm{q}}=-J_{\bm{q}+\bm{Q}}$.

The polarization functions $\Pi_{\rm A}^{\lambda}$ and $\Pi_{\rm B}^{\lambda}$ are computed separately.
Namely, 
we have two effective impurity models corresponding to A and B sublattices.
The effective interaction ${\cal J}_{\alpha}^{\lambda}$ for each impurity model is determined so that the corresponding local susceptibility 
$\chi_{\rm loc(\alpha)}^{\lambda}(i\nu_n)$ 
equals to the site-diagonal susceptibility in Eq.~(\ref{eq:chi_diag-AFM}): 
$\chi_{\rm loc(\alpha)}^{\lambda}(i\nu_n) = \chi_{\alpha}^{\lambda}(i\nu_n)$.

Without magnetic field,
we have 
$\langle S_{\rm B} \rangle = -\langle S_{\rm A} \rangle$
and
$\Pi_{\rm B}^{\lambda}(i\nu_n) = \Pi_{\rm A}^{\lambda}(i\nu_n)^*$.
Hence, 
only one impurity problem needs to be solved in this case.

\subsection{Ising limit}
\label{sec:ising}
In the Ising (classical) limit,
all the dynamical quantities are replaced by their static values, e.g., 
$\Pi(i\nu_n) = \delta_{n0} \Pi$ and 
${\cal J}(i\nu_n) = \delta_{n0} {\cal J}$.
Hence, the effective local model consists only of static fields, and can be solved explicitly.
The magnetization 
$m=2\langle S^z \rangle$ for the FM state is given by
\begin{align}
m = \tanh [ \beta (zJ - {\cal J})m].
\label{eq:ising1}
\end{align}
The difference from the MF approximation for the Ising model
is the presence of the reaction field ${\cal J}$,
which is determined from the self-consistency condition (\ref{eq:self_consist}).
Eliminating $\Pi$, we obtain
\begin{align}
\frac{1}{N} \sum_{\bm{q}}
\frac{1}{1-\chi_{\rm loc} (J_{\bm{q}}-{\cal J})} = 1.
\label{eq:ising2}
\end{align}
The local static susceptibility $\chi_{\rm loc}$ is given in terms of $m$ by
$\chi_{\rm loc} = (1-m^2)/4T.$
The above equations correspond to the spherical model approximation by Brout.\cite{Brout60}

Equations for the AFM state are also reduced to the above equations 
by interpreting $m$ as the staggered magnetization.

\section{Numerical results}
\label{sec:results}
In this section, we present the numerical solution of the self-consistent equations in dimensions $2 \leq d \leq 5$.
We have solved the effective local problem by using CT-QMC applied to models with the spin-boson coupling.\cite{Otsuki-SB, Gull11, note-CTQMC}
We set the transition temperature $T_{\rm c}^{\rm MF}$ in the mean-field approximation to unity, i.e., 
$T_{\rm c}^{\rm MF}=z|J|/4=1$.

\subsection{Ising limit}
\label{app:ising}

\begin{figure}[t]
	\begin{center}
	\includegraphics[width=\linewidth]{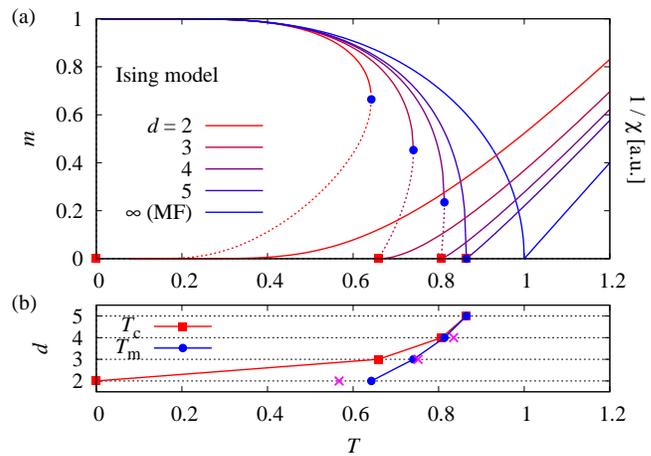}
	\end{center}
	\caption{(Color online) (a) The ordered moment $m$ and the inverse of the static susceptibility for the FM Ising model in dimensions $2 \leq d \leq 5$. 
	The dotted lines show $m$ for the unstable solution.
	(b) Transition temperatures.
	The red squares, indicated by $T_{\rm c}$, correspond to divergence of $\chi_{\rm F}$, and the blue circles, $T_{\rm m}$, correspond to  vanishing $m$.
	The crosses indicate the exact result for $d=2$, and Monte Carlo results for
$d=3$ from Ref.\cite{Ferrenberg-Landau91} 
and for $d=4$ from Ref.\cite{Lundow-Markstroem09}.}
	\label{fig:ising}
\end{figure}

We first present results in the Ising limit in order to establish the critical properties within the s-DMFT.
Figure~\ref{fig:ising}(a) shows the inverse of the uniform susceptibility $\chi$ in the PM phase and the ordered moment $m$ in the FM phase.
The transition temperature $T_{\rm c}$ is evaluated from the divergence of $\chi$ as
$T_{\rm c}=0$ ($d=2$),
0.659 ($d=3$),
0.807 ($d=4$), and 
0.865 ($d=5$).
The critical exponent $\gamma$ defined by $\chi \propto (T-T_{\rm c})^{-\gamma}$ increases with decreasing $d$.
From analytical calculations (see Appendix~\ref{sec:exponent}), 
we obtain $\gamma=2$ in $d=3$,
which is about 60\% larger than the Monte Carlo result, $\gamma=1.239$.\cite{Ferrenberg-Landau91}
For $d=4$, $\chi$ does not show the power-law behavior.
For $d > 4$, the exponent takes the mean-field value, $\gamma=1$, 
which is consistent with the exact upper critical dimension $d_{\rm c}=4$.\cite{Aizenman81}
The moment $m$ for the stable solution shows discontinuity in $d \leq 4$, indicating the first-order transition.
This is an artifact of the present approximation, and remains also in the quantum case as discussed later.

Figure~\ref{fig:ising}(b) shows the transition temperatures $T_{\rm c}$ evaluated from the divergence of $\chi$, and $T_{\rm m}$ evaluated from vanishing $m$.
The exact and Monte Carlo results are also plotted for comparison.
The difference between $T_{\rm c}$ and $T_{\rm m}$ becomes smaller as $d$ increases, and 
eventually vanishes at $d=5$,
where $m$ changes continuously and the critical exponent $\gamma$ is reduced to the MF value, $\gamma=1$.

\subsection{Static properties}

\begin{figure}[tb]
	\begin{center}
	\includegraphics[width=\linewidth]{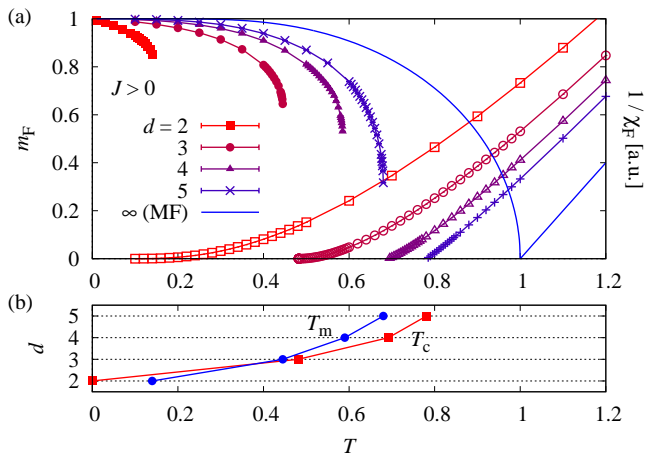}
	\end{center}
	\caption{(Color online) (a) The FM moment $m_{\rm F}$ and the inverse of the static uniform susceptibility $\chi_{\rm F}$ as a function of $T$ for $J>0$.
	(b) Transition temperatures (see Fig.~\ref{fig:ising} for the definition of $T_{\rm c}$ and $T_{\rm m}$).}
	\label{fig:mag-FM}
\end{figure}

We proceed to the quantum system, first with the FM interaction, $J>0$.
Figure~\ref{fig:mag-FM}(a) shows the inverse of the static uniform susceptibility $\chi_{\rm F}$ in the PM phase and the magnetic moment $m_{\rm F}=2 \langle S^z_i \rangle$ in the FM phase.
The transition temperature $T_{\rm c}$ evaluated from the divergence of $\chi_{\rm F}$ is
$T_{\rm c}=0$ ($d=2$),
0.481 ($d=3$),
0.692 ($d=4$), and
0.781 ($d=5$).
The critical exponent $\gamma$ is the same as in the Ising limit,
as checked numerically. However
the ordered moment $m_{\rm F}$ shows discontinuity even in $d=5$ 
in contrast to the Ising limit.
As $T$ decreases, $m_{\rm F}$ approaches 1 for all dimensions.
This follows from the fact that the full-moment state is an eigenstate of the Hamiltonian (\ref{eq:hamil}).
Indeed, the influence of the fluctuations vanishes as $T \to 0$,
as can be estimated as the average of the expansion order contributing in the CT-QMC.

Figure~\ref{fig:mag-FM}(b) shows transition temperatures $T_{\rm c}$ and $T_{\rm m}$.
Both quantities approach 1 in the MF limit $d \to \infty$.
In $d>2$, it shows an unphysical situation $T_{\rm c} > T_{\rm m}$. 
This  means that no stable solution exists in the temperature range $T_{\rm m} < T < T_{\rm c}$.
Actually, the present equations give a stable PM solution even in this range.
This situation is related to the fact that the present theory does not satisfy the Ward identity,
which will be discussed in more detail in the next section.

\begin{figure}[tb]
	\begin{center}
	\includegraphics[width=\linewidth]{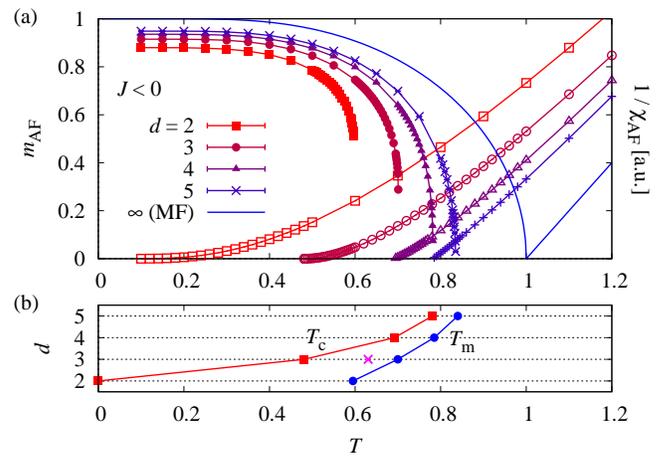}
	\end{center}
	\caption{(Color online) (a) The AFM moment $m_{\rm AF}$ and the inverse of the static staggered susceptibility $\chi_{\rm AF}$ for $J<0$.
	(b) The transition temperatures (see Fig.~\ref{fig:ising} for the definition of $T_{\rm c}$ and $T_{\rm m}$). The cross point indicates the QMC result.\cite{Sandvik98}}
	\label{fig:mag-AFM}
\end{figure}

We next show results for the AFM interaction, $J<0$, in Fig.~\ref{fig:mag-AFM}.
The staggered susceptibility $\chi_{\rm AF}$ in this case is equivalent to $\chi_{\rm F}$ in $J>0$.
The AFM moment 
$m_{\rm AF}=2 \langle S^z_{\rm A} \rangle =-2 \langle S^z_{\rm B} \rangle$
at low enough temperatures
decreases as $d$ decreases.
The value $m_{\rm AF}=0.915$ obtained in $d=3$ is larger than the value 0.844 evaluated in the spin-wave theory.
Similarly in $d=2$, we obtain $m_{\rm AF}=0.880$, while the stochastic series expansion gives 0.614.\cite{Sandvik97}
These comparisons indicate that account of the mixing between spins on A and B sublattices is insufficient in our approximation.
The transition is of first-order as in the FM case,
while the condition $T_{\rm m} > T_{\rm c}$ is satisfied in this case as shown in Fig.~\ref{fig:mag-AFM}(b).
Although $m_{\rm AF}$ exhibits only a small discontinuity $\Delta m_{\rm AF} \lesssim 0.03$ in $d=5$, the transition temperatures $T_{\rm m}$ and $T_{\rm c}$ do not coincide.
This inconsistency is related again to violation of the Ward identity as in the unphysical situation of the FM solution.

\subsection{Excitation spectrum}
\begin{figure}[tb]
	\begin{center}
	\includegraphics[width=\linewidth]{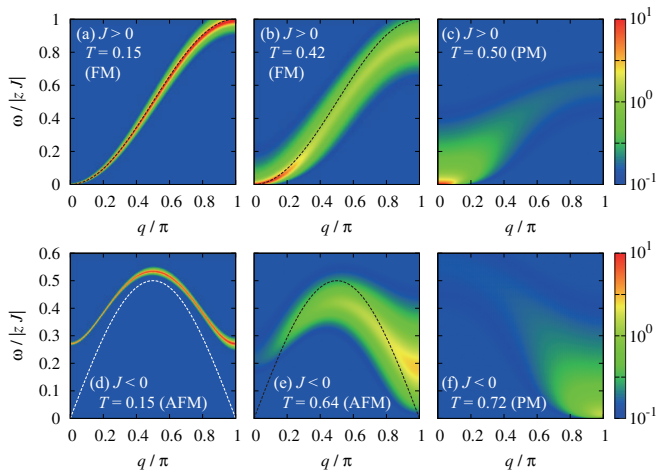}
	\end{center}
	\caption{(Color online) Intensity plots of the magnetic excitation spectrum ${\rm Im}\chi^{\perp}_{\bm{q}} (\omega+i0)$ in $d=3$ for 
(a)--(c) ferromagnetic ($J>0$), and
(d)--(f) antiferromagnetic ($J<0$) cases.
The colors indicate the intensity in a logarithmic scale. The momentum $\bm{q}$ runs along $\bm{q}=(q, q, q)$. The dashed line shows the energy dispersion of the magnon evaluated by the spin-wave theory. 
}
	\label{fig:spec}
\end{figure}

We study the transverse fluctuations $\chi^{\perp}_{\bm{q}}$ 
in the ordered phase as well as in the PM phase by
analytical continuations $i\nu_n \to \omega +i0$ 
using the Pad\'e approximation. 
The spectra 
${\rm Im}\chi_{\bm{q}}^{\perp}(\omega+i0)$
in $d=3$
are shown 
in Fig.~\ref{fig:spec}(a)--(c) for the FM interaction, $J>0$, and 
in Fig.~\ref{fig:spec}(d)--(f) for the AFM interaction, $J<0$.
The intensity is plotted in a logarithmic scale so that a tail of the peak is visible.
The momentum $\bm{q}$ runs along the line $\bm{q}=(q, q, q)$, on which $\gamma_{\bm{q}}$ simply becomes $\gamma_{\bm{q}}=\cos q$.
We note that
the spectrum is plotted in the original Brillouin zone even in the AFM phase without its folding.

In the PM phase near the transition temperature [Fig.~\ref{fig:spec}(c) and (f)],
low-energy excitations appear around $\bm{q}=\bm{0}$ or $\bm{Q}$, reflecting strong fluctuations close to the magnetic ordering.
This paramagnon spectra 
demonstrate the inclusion of the fluctuations in the present theory, 
going beyond the mean-field solution.
In the FM phase [Fig.~\ref{fig:spec}(a) and (b)], 
the spectra exhibits magnon excitations.
At low temperatures, the energy spectrum agrees well with that calculated by the spin-wave theory.
The broadening 
near the transition temperature has also been obtained in the present theory.
It corresponds to the effect of interactions between thermally excited magnons in the spin-wave description.
In the AFM phase [Fig.~\ref{fig:spec}(d) and (e)], 
on the other hand, 
it turns out that the magnon spectrum
shows a fictitious energy gap.
Namely, the present approximation fails to reproduce gapless magnon excitations in the AFM phase.

\section{Discussion}
\label{sec:discussion}

\begin{figure}[tb]
	\begin{center}
	\includegraphics[width=0.90\linewidth]{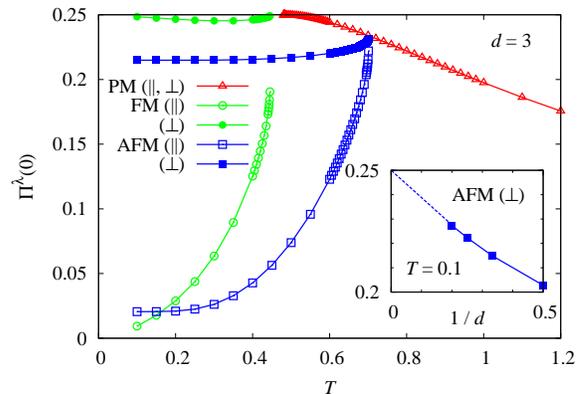}
	\end{center}
	\caption{(Color online) The static part of the polarization function $\Pi^{\lambda}(0)$ for $d=3$ as a function of $T$. 
Starting from the PM phase,	the condition $\Pi^{\lambda}(0)=1/4$ 
gives divergence of the susceptibility, while in the ordered phase,
$\Pi^{\perp}(0)=1/4$ is necessary for gapless excitations. 
	The inset shows $\Pi^{\perp}(0)$ in the AFM phase at $T=0.1$ as a function of $1/d$.}
	\label{fig:Pi}
\end{figure}

In this section, we discuss the origin of the fictitious energy gap of AFM magnons.
We first analyze the low-energy excitations of the spectra in Fig.~\ref{fig:spec}.
From the expressions for $\chi_{\bm{q}}(i\nu_n)$ in Eqs.~(\ref{eq:chi_q}) and (\ref{eq:chi_q-AFM}), 
we obtain the condition for gapless excitations as
\begin{align}
{\rm Re}\Pi_{\alpha}^{\perp}(0)
= J_{\bm{q}_0}^{-1}.
\label{eq:gapless}
\end{align}
Here, $\bm{q}_0$ is the ordering vector, i.e. 
$\bm{q}_0=\bm{0}$ for $J>0$ and
$\bm{q}_0=\bm{Q}$ for $J<0$.
The right-hand side corresponds to $J_{\bm{q}_0}^{-1} =1/4$ in the present energy unit.
The condition above is equivalent to 
divergence of the static susceptibility in the PM phase.
Figure~\ref{fig:Pi} shows the temperature dependence of ${\rm Re}\Pi^{\lambda}(0)$ in $d=3$.
In the PM phase, $\Pi^{\perp}(0)=\Pi^{\parallel}(0)$ reaches at 1/4 at $T=T_{\rm c}$. 
The AFM solution begins to
split from the PM solution already at $T=T_{\rm m}>T_{\rm c}$.
Hence, the resultant first-order transition causes
deviation of $\Pi^{\perp}(0)$ from condition~(\ref{eq:gapless}).
To see the dependence on spatial dimensions,
we plot $\Pi^{\perp}(0)$ at $T=0.1$ as a function of $1/d$ in the inset of Fig.~\ref{fig:Pi}.
This value reasonably extrapolates to 1/4 as $d\to \infty$, meaning that the fictitious energy gap closes in the limit $d\to \infty$. 

As is well known, the gapless magnon excitations can be described by variants of the random phase approximation (RPA) such as the spin-wave theory.
To see the form of $\Pi^{\perp}(i\nu_n)$,
we evaluate it in the decoupling approximation.\cite{Kawasaki-Mori62, Tahir-Kheli62, Callen63}
This approximation, at low temperatures, provides the same results with the spin-wave theory,
and further includes thermal effects through $\langle S^z \rangle$.
Deriving the equation of motion for $\chi_{\alpha}^{\perp}(i\nu_n)$,
and applying
the decoupling
$\langle S_i^z S_j^+; S_k^- \rangle \simeq \langle S_i^z \rangle \langle S_j^+; S_k^- \rangle$,
we obtain 
\begin{align} 
\Pi_{\alpha}^{\perp}(i\nu_n)
=
\frac{-\langle S^z_{\alpha} \rangle}
{i\nu_n 
-zJ
\langle S^z_{-\alpha} \rangle}.
\label{eq:Pi-RPA}
\end{align}
Here, the second term in the denominator is due to the MF.
It is noteworthy  that $\Pi_{\alpha}^{\perp}$ is local as in the s-DMFT.
This expression shows the correct high-frequency behavior
$\Pi_{\alpha}^{\perp}(i\nu_n) \sim -\langle S^z_{\alpha} \rangle / i\nu_n$.
From this expression, we obtain $\Pi_{\alpha}^{\perp}(0)=|zJ|^{-1}$.
Namely the decoupling approximation satisfies condition (\ref{eq:gapless}) as expected.

The present formalism of the s-DMFT satisfies the stationary condition of the thermodynamic potential but not the Ward identity.\cite{Kuramoto-Fukushima98}
Namely, consistency between thermodynamic quantities and fluctuations is not ensured.
This could result in an inconsistency between the PM solution and the magnetic solution.
To check the Ward identity, we have computed $\chi_{\rm F}$ and $\partial m_{\rm F}/\partial h$ under magnetic field $h$ ($\chi_{\rm AF}$ and $\partial m_{\rm AF}/\partial h_{\bm{Q}}$ for $J<0$) [figure not shown].
As a result, we found that they do not agree with each other 
for all values of $h$ in the Heisenberg model.\cite{note-WI}
Furthermore, for $J>0$ we have observed a first-order transition for both $h \neq 0$ and 
$h=0$.
We thus conclude that the breakdown of the Ward identity is responsible for the fictitious first-order transition, and hence for the energy gap of the magnetic excitations.
In order to satisfy the Ward identity {\em and} include fluctuations of order $1/z$, the momentum dependence of $\Pi_{\bm{q}}(i\nu_n)$ needs to be taken into account.\cite{Kuramoto-Fukushima98}
Constructing such a theory requires much more 
elaborate consideration even in the Ising limit.\cite{Englert63}

Although the s-DMFT describes FM magnons reasonably well, the Nambu-Goldstone theorem nevertheless is not satisfied exactly. 
In fact, slight deviation from condition~(\ref{eq:gapless}) can be seen in Fig.~\ref{fig:Pi} 
in the middle temperature range.
The low-temperature gapless excitations in the FM case are due to vanishing fluctuations. 
Indeed, our numerical solution for $\Pi^{\perp}(i\nu_n)$ agrees with the RPA result in Eq.~(\ref{eq:Pi-RPA}) at low temperatures.
On the other hand, the gapless feature near $T_{\rm m}$ comes from the (accidentally) fortunate situation $T_{\rm m}\lesssim T_{\rm c}$ as shown in Fig.~\ref{fig:mag-FM}.
On the contrary, $T_{\rm m}$ in the AFM case 
is much larger than $T_{\rm c}$ 
as shown in Fig.~\ref{fig:mag-AFM}.

Finally, we discuss condition (\ref{eq:gapless}) in analogy with 
the ordinary (fermionic) DMFT.
In the s-DMFT,
$\Pi(\omega)$ may be regarded as a self-energy of the auxiliary bosonic field which mediates the exchange interaction.\cite{Kuramoto-Fukushima98}
Hence, we have a correspondence between $\Pi(\omega)$ in the s-DMFT and the electron self-energy $\Sigma(\omega)$ in the DMFT.
The real part of $\Sigma(\omega)$ is responsible for the shift of the Fermi energy,
and the Luttinger theorem on the Fermi-surface volume demands conservation of the effective chemical potential 
$\mu_{\rm eff} = \mu - {\rm Re}\Sigma(0)$.\cite{Georges96}
From numerical calculations for some models, it has been demonstrated that the DMFT solution satisfies this condition.\cite{Otsuki-LFS, Georges96}
By analogy, we could expect condition (\ref{eq:gapless}) to be satisfied in the s-DMFT as well, but our result was opposite of the expectation.
Note that the bosonic counterpart of the Luttinger theorem is lacking.

\section{Summary}
\label{sec:summary}
We have presented the numerical solution of the self-consistent equations which may be regarded as a spin version of the DMFT.
The theory incorporates dynamical fluctuations around the MF within the local approximation.
We have demonstrated that 
fluctuations in this theory lead to paramagnon excitations near the transition temperature.
Furthermore, the magnon spectrum in the FM phase behaves reasonably.
In the AFM phase, on the other hand, a fictitious energy gap appears in the magnon spectrum.
This artifact is related to the first-order transition in the present approximation and is ascribed to 
breakdown of the Ward identity.\cite{Akerlund13, footnote}

It is an interesting open problem how to recover the gapless magnon mode
in the AFM phase.
Our analysis has made it clear that one should at least include a momentum dependence of the polarization function (or the bosonic self-energy). 
For this purpose, one may proceed to an extension that replaces the effective single-site problem with a cluster impurity problem,\cite{Maier05}
or to another extension that 
uses the perturbative expansion around the s-DMFT and the extended DMFT.\cite{Rubtsov12}

\begin{acknowledgments}
We thank M. Potthoff for useful suggestion on the unstable solution in the first-order transition.
One of the authors (J.O.) is supported by JSPS Postdoctoral Fellowships for Research Abroad.
\end{acknowledgments}

\appendix
\section{The effective interaction in ferromagnetic phase}
\label{app:J_eff}
In this appendix, we derive the effective interaction ${\cal J}^{\perp}(i\nu_n)$ in Eq.~(\ref{eq:J_eff2}) from the Hamiltonian $H_{\rm imp}^{\rm FM}$ in Eq.~(\ref{eq:H_imp-FM}).
The effective interaction ${\cal J}^{\perp}(i\nu_n)$ is given by the correlation function
\begin{align}
{\cal J}^{\perp}(i\nu_n)
= g_{\perp}^2
\langle \alpha \phi^{+} - i\beta \pi^{+};
\alpha \phi^{-} + i\beta \pi^{-} \rangle_0,
\label{eq:J_eff-app}
\end{align}
in the Matsubara-frequency domain,
where the subscript 0 indicates an average in the noninteracting system, $g_{z}=g_{\perp}=0$.
These correlations consist of the bosonic Green functions
$G_{\pm}(i\nu_n) = -\langle b_{\pm} ; b_{\pm}^{\dag} \rangle_0$.
The explicit expressions are given by
\begin{align}
G_{\pm}(i\nu_n)
=\frac{1}{N} \sum_{\bm{q}} \frac{1}{i\nu_n-\omega_{\bm{q}\pm}}.
\end{align}
The definition of $\phi^{\pm}$ and $\pi^{\pm}$ in Eq.~(\ref{eq:op_perp}) leads to the relations
\begin{align}
&\langle \phi^+; \phi^- \rangle_0
=\langle \pi^+; \pi^- \rangle_0
= -[G_{+}(i\nu_n) + G_{-}(-i\nu_n)],
\nonumber \\
&\langle \phi^{+}; \pi^{-} \rangle_0
= -\langle \pi^+; \phi^- \rangle_0
= -i[G_{+}(i\nu_n) - G_{-}(-i\nu_n)].
\end{align}
Inserting these expressions into Eq.~(\ref{eq:J_eff-app}),
and using the parameterization for $\alpha$ and $\beta$,
we obtain Eq.~(\ref{eq:J_eff2}).

\section{Critical exponent}
\label{sec:exponent}
In this appendix, we derive the critical exponent of the susceptibility by analytical calculations in the Ising limit.
Equation~(\ref{eq:ising2}) may be written as
$zJ \chi_{\rm loc} = F(1/zJ\Pi)$
where 
\begin{align}
F(x) = \frac{1}{N} \sum_{\bm{q}} \frac{1}{x-\gamma_{\bm{q}}}.
\end{align}
The uniform susceptibility $zJ\chi=[(zJ\Pi)^{-1}-1]^{-1}$
diverges at the point where $zJ\Pi=1$ is satisfied. 
Inserting $\chi_{\rm loc}=1/4T$ and using $T_{\rm c}^{\rm MF}=zJ/4$, 
we obtain the expression for the 
critical temperature as 
$T_{\rm c} / T_{\rm c}^{\rm MF} = 1/F(1)$.
The behavior of $F(x)$ around $x=1$ yields the critical properties of $\chi$. 
For each dimension, $F(x)$ is evaluated around $x=1$ as
\begin{align}
F(1+y) \sim 
\begin{cases}
\displaystyle{-(1/\pi) \ln 4y} &(d=2) \\
\displaystyle{F(1) -(\sqrt{27}/4 \pi) \sqrt{y}} &(d=3) \\
\displaystyle{F(1) + (1/\pi^2) y \ln y} &(d=4) \\
F(1) -a y &(d \geq 5)
\end{cases}.
\end{align}
For $d=2$, $F(x)$ diverges at $x=1$, 
leading to absence of the second-order phase transition at finite temperatures.
In higher dimensions, on the other hand, $F(1)$ is finite:
$F(1)=1.531$ ($d=3$),
1.240 ($d=4$), and 
1.156 ($d=5$).
The $\sqrt{y}$ term in $d=3$ and the $y$-linear term in $d=5$ give rise to the critical exponent $\gamma=2$ and $\gamma=1$, respectively.


\begin{thebibliography}{99}
\bibitem{note-CTQMC} The Hamiltonian~(\ref{eq:H_imp-FM}) for the magnetic phase is different from that written in Ref.~\cite{Otsuki-SB}. However, since their actions are equivalent, the algorithm can be applied to the present model without any modifications. We may take care of only the discontinuity in ${\cal J}^{\perp}(\tau)$ at $\tau=0$ corresponding to the slow decay in the frequency domain.
\bibitem{note-WI} In the Ising limit, the Ward identity is satisfied only in the absence of the magnetic field.\cite{Kuramoto-Fukushima98}
\bibitem{footnote} 
Very recently, the DMFT has been applied to a $\varphi^4$ quantum field theory which exhibits a phase transition as in the Heisenberg model.\cite{Akerlund13} Their result shows a first-order transition in $d=2, 3$ similarly to our results.
\end{thebibliography}
\end{document}